\newif\ifsubmode
\submodefalse
\submodetrue

\documentclass[iop,apj]{emulateapj}   

\usepackage{natbib}
\usepackage{subfigure}
\usepackage{xspace}   
\usepackage{color}   
\usepackage{epsfig}
\usepackage{graphicx}

\newcommand{\jw}{\emph{JWST}\xspace}
\newcommand{\crr}{CR\xspace}
\newcommand{\crs}{CRs\xspace}
\newcommand{\dif}{\textsc{2pt diff}\xspace}
\newcommand{\dev}{\textsc{dev from fit}\xspace}
\newcommand{\yint}{\textsc{y-int}\xspace}
\newcommand{\pn}{photon-noise\xspace}
\newcommand{\rn}{read-noise\xspace}

\shorttitle{Detecting Cosmic Rays in Infrared Data} 
\shortauthors{Anderson \& Gordon}

\begin{document}

\title{Optimal Cosmic-Ray Detection for Nondestructive Read Ramps} 

\author{Rachel E. Anderson and Karl D. Gordon} \affil{Space Telescope
  Science Institute, 3700 San Martin Drive, Baltimore, MD 21218}
\email{randers@stsci.edu, kgordon@stsci.edu}

\date{\today}


\begin{abstract}
  Cosmic rays are a known problem in astronomy, causing both loss of
  data and data inaccuracy.  The problem becomes even more extreme
  when considering data from a high-radiation environment, such as in
  orbit around Earth or outside the Earth's magnetic field altogether,
  unprotected, as will be the case for the \emph{James Webb Space
    Telescope}~(\jw).  For \jw, all the instruments employ
  nondestructive readout schemes.  The most common of these will be
  ``up the ramp'' sampling, where the detector is read out regularly
  during the ramp.  We study three methods to correct for cosmic rays
  in these ramps: a two-point difference method, a deviation from the
  fit method, and a y-intercept method.  We apply these methods to
  simulated nondestructive read ramps with single-sample groups, and
  varying combinations of flux, number of samples, number of cosmic
  rays, cosmic-ray location in the exposure, and cosmic-ray strength.
  We show that the y-intercept method is the optimal detection method
  in the \rn-dominated regime, while both the y-intercept method and
  the two-point difference method are best in the
  \pn-dominated regime, with the latter requiring fewer computations.
\end{abstract}



\section{Introduction}\label{s:intro}

Advances in technology that allow us to observe fainter objects, build
more complex systems, and send telescopes further into space have
challenged us to continue to improve our calibrations.  This includes
detection methods for cosmic rays (\crs) or any charged particle that
adds a jump to the data.  \jw's orbit at the second Earth-Sun Lagrange
point (L2), which allows passive cooling of the telescope to $\sim $50
K, puts it outside the protective mantel of the Earth's magnetic
field.  This could make \crs a serious problem.  Furthermore, newer
infrared telescopes will have a lower \rn, and thus we will detect
lower \crs.  Finally, long observing times will be necessary to
complete many of the scientific goals of the \jw, causing \crs to be
an even larger problem.  From a study by~\citet{Rob09_3} we can
calculate that in every 2000 s, on average, 13\% of the pixels on the
\jw HgCdTe detectors and 25\% of the pixels on the \jw Si:As detectors
will be affected by \crs.  These values could be even greater, since
this study does not take into account secondary particles.  For
comparison, onboard measurements of the \emph{NICMOS} camera on the
\emph{HST} show that about 10\% of the pixels show a \crr hit for
every 2000 s of integration~\citep{Via09}.  These \crs will
include low-energy \crs from secondary particles and the Sun, and
higher-energy galactic \crs~\citep{Rob09}.  Clearly, a reliable method
to detect both low- and high-energy \crs is needed.



In this article we will discuss \crr detection methods for infrared
data, which use nondestructive read ramps.  \citet{Off01} found that
in order to correct for \crs, the nondestructive readout scheme was
most efficient.  By integrating the charge on each pixel in this way
it is possible to calculate the slope of these ramps in counts per time
to get the flux of the sky (Rieke 2007).  We refer to the integrated
charge as a ramp made up of a specified number of samples.  We discuss
the slope of these ramps as the calculation of the flux.  Finally, a
\crr affects the ramp between samples, however we use `sample
number' to refer to the first sample after the \crr hit.

Correcting for \crs in ramps is not a new problem~\citep{Off99}.
However, the advent of large nondestructive arrays in orbit coupled
with ground-based processing of the ramps means, when it comes to
calculating the slope of ramps, that there are more options available
for \crr detection.  In addition to \crs, noise is added to the data
by the detector and readout electronics as well (Fanson 1998, Tian et
al. 1996, Rieke 2007).  In this article we focus on \pn and \rn for
our correlated and uncorrelated noise, respectively; however, we are not
restricted to just these two terms.  The work in this article is general
and can account for any correlated and uncorrelated noise sources.
Therefore, our question is, What is the best we can do at finding \crs
in a ramp, given the noise in the ramp?

To understand and test various \crr detection methods for ramps, we
simulate infrared data as described in Section~\ref{s:sim}.  Anytime a
slope is calculated for the preceding process or for the \crr
detection methods, we do so using linear regression for data with
correlated and random uncertainties (in consideration of the \pn and
the \rn, respectively), as described in Section~\ref{s:corr}.  In
Section~\ref{s:methods} we propose three techniques to detect \crs.
The first is a two-point difference method, the second is a deviation
from the fit method, and the third is a y-intercept comparison.  There
are various conditions that can hinder/aid in \crr rejection
(e.g., number of samples, slope, number of \crs, size of \crs, and
location of \crs); therefore, we aim to study combinations of these
and to find which algorithm behaves the best under different conditions.
These results are presented in Section~\ref{s:results}.  A discussion
on our findings is in Section~\ref{s:dis}.  Concluding remarks can be
found in Section~\ref{s:cons}.

\section{Simulating Nondestructive Read Ramps}
\label{s:sim}
In order to test how well a \crr detection method works, we have to
test the method on a known \crr.  For that reason, we have simulated
ramps with \pn and \rn added in, to which we can either add \crs with
known location in the exposure and known amplitudes or leave the ramps
clean (\crr-free).  We build these ramps with the slope, y-intercept,
number of samples, time between samples (sample time), and \rn as
inputs, using parameters given in Table~\ref{tbl:params} as guidelines
(slope and number of samples will be changed in this study).  The only
value that is instrument-specific is the \rn, which was chosen to
match the expected value of the Mid-Infrared Instrument (\emph{MIRI})
on the \jw.  Note that we do not group any of the samples by taking a
weighted average or coadding.  Furthermore, we assume uniform sampling
(i.e., constant time between samples), as that is what is used by the
\jw and also so that calculations are more efficient.  We also assume
that a nonlinearity correction has already been performed with no
error, and we do not account for the effects of quantization noise.
Although we are simulating \emph{MIRI} detector parameters, this is
only an example.  The methods discussed in this article will apply to
any other nondestructive read data, including those for the \jw
near-infrared detectors (\emph{Near-Infrared Camera} [\emph{NIRCam}]
and \emph{Near-Infrared Spectrograph} [\emph{NIRSpec}]) the \emph{HST}
infrared detectors (\emph{Wide Field Camera 3 - IR} [\emph{WFC3-IR}]
and \emph{Near-Infrared Camera and Multi-Object Spectrometer}
[\emph{NICMOS}]) and ground-based detectors, by simply changing the
values in Table~\ref{tbl:params}.  If the data you would like to
simulate do included grouped samples, however, some revisions will
have to be made.

\begin{table}
  \begin{center}
    \begin{tabular}{ll}
      \hline
      \hline
      Parameters & Values \\
      \hline
       Slope & 70.0 $e^-$/s \\
       Y-Intercept & 21000.0 $e^-$ \\
       Number of Samples & 40 \\
       Sample Time & 27.7 s \\
       Read Noise & 16.0/$\sqrt{8}$ $e^-$/sample \\
       \hline
     \end{tabular}
     \caption[MIRI Parameters]{Parameters Used to Create MIRI Ramps.
       \label{tbl:params}}
  \end{center}
\end{table}

At time $t=0$ the counts $y_0$ are equal to the
y-intercept.\footnote{Although we use the value for the y-intercept
  given in Table~\ref{tbl:params}, this is an arbitrary value and does
  not change our calculations.}  This is not considered part of the
ramp, because $t=0$ is when the reset occurred, but we start reading at
time equal to one sample time, $t_1$.

Each signal is comprised of the counts from the previous signal plus
the additional signal, the \pn from the additional signal, and the
\rn.  To construct the ramps, we follow this procedure:

\begin{enumerate}
\item For each ramp calculate the expected signal (i.e., no noise,
  clean), $s_e$, defined as
\begin{eqnarray}
 s_e \equiv m  t_s,
\label{eq:se}
\end{eqnarray} 
where $m$ is the input slope, and $t_s$ is the sample time.  
\item For each sample calculate the sum of the expected signal and the
  unique \pn of this signal ($s_e + \sigma_{n_pi}$) from a Poisson
  distribution with $\lambda = s_e$.  Add this to the signal from the
  previous sample (or y-intercept if it is the first sample).  This
  step is shown in equation~\ref{eq:y0}.  Since we have only added the
  correlated noise, and we still need to add the uncorrelated noise,
  we have called the signal $y_i^*$ instead of $y_i$).
\begin{eqnarray}
y_i^* = y_{i-1}^* + s_e + \sigma_{n_pi}.  
\label{eq:y0}
\end{eqnarray} 
\item If there are to be one or more \crs to the ramp, simply add the
electrons with the expected signal:
\begin{eqnarray}
y_i^* = y_{i-1}^* + s_e + \sigma_{n_pi} + CR_{mag},  
\end{eqnarray} 
where $CR_{mag}$ is the magnitude of the \crr.
\item Read noise, denoted by $n_r$, is due to readout electronics;
  consequently, it is uncorrelated.  Therefore, when all samples are
  populated, add a unique \rn, $\sigma_{n_ri}$, to each sample
  (equation~\ref{eq:y1}):
\begin{eqnarray}
 y_i = y_i^* + \sigma_{n_ri}.  
\label{eq:y1}
\end{eqnarray} 
For each sample $\sigma_{n_ri}$ is taken from a Gaussian with $n_r$ as
the standard deviation.
\end{enumerate}

The uncertainties for the samples in the ramp are calculated by adding
the \pn and \rn in quadrature:
\begin{eqnarray}
 \sigma_{y} = \sqrt{n_r^2 + n_p^2},
\label{eq:yerr}
\end{eqnarray} 
where $n_p$ is the Poisson noise of the expected signal in the ramp;
thus, $n_p = \sqrt{mt_s}$, and $n_r$ is taken from
Table~\ref{tbl:params}.  We have chosen to give each sample equal
weighting, rather than choosing the \pn to be the Poisson noise of the
counts in the ramp, so that we can improve our results by avoiding a
bias based on sample number.

\section{Linear Fit Algorithm for Data with Correlated
  Errors}\label{s:corr}
For two of the three \crr detection methods, it is very important that
we use the best calculation of the slope and y-intercept. Therefore we
must use all information including correlated and uncorrelated errors.
\citet{Gor05} explains how to take into account both correlated and
uncorrelated errors for the slope and y-intercept uncertainties, and
here we describe how to include this information for the fit as well,
using matrix notation and specifically making use of the covariance
matrix.  For a refresher on calculating a linear fit using matrices as
well as the use of the covariance matrix, see \citet{Hog10}.  Every
time a slope is calculated in the CR detection methods described in
this article, we do so using this method.

The real power of using matrix notation comes from the covariance
matrix, $\bf{C}$, as it includes information about the degree of
correlation between samples.  Furthermore, the slope and y-intercept
uncertainties resulting from using this covariance matrix
automatically take into account the correlated and uncorrelated
errors, provided they are included in the covariance matrix.  $\bf{C}$
is defined as:
\begin{eqnarray}
\bf{C} =
\left[ {\begin{array}{cccc}
 \sigma^2_{y_1} & c_{1,2} & ... & c_{1,N} \\
 c_{2,1} & \sigma^2_{y_2} & ... & c_{2,N} \\
 ... & ... & ... & ...\\
 c_{N,1} & c_{N,2} & ... & \sigma^2_{y_N}  \\
 \end{array} } \right].
\end{eqnarray}

The values on the diagonal of $\bf{C}$ are the uncertainties of the
$y_i$, whereas the off-diagonal elements are the covariance between
the different $y_i$.  If the data are uncorrelated, then the
off-diagonal matrix elements are all zero, $c_{i,j} = 0, i \neq j$.
However, we do have correlated data due to the \pn.

When constructing $\bf{C}$ for correlated errors, we follow the logic
of~\citet{Fix00} and think of $\bf{C}$ as the sum of two matrices:
one for the \pn, $\bf{P}$, and one for the \rn, $\bf{R}$.
\citet{Fix00} define these matrices for the case where the data are
the two-point difference of the samples (the \pn is not correlated,
and the \rn is correlated).  However, for fitting a line to the individual
samples, it is just the opposite.  The \rn is uncorrelated, and thus
$\bf{R}$ is just $n_r^2$ on the diagonal and zero elsewhere.  

The diagonal of $\bf{P}$ is $n_p^2$, just as it would be for
uncorrelated data, but since the \pn is correlated, the off-diagonal
elements are not zero.  An estimate of all of the \pn in $y_i$ is the
\pn added to each sample multiplied by $i$ (i.e., $i n_p$).  The
correlation between samples $y_i$ and $y_j$ is the \pn that is in
$y_i$ that is also in $y_j$.  The elements in $\bf{P}$ that represent
this correlation are $p_{i,j}$ and $p_{j,i}$.  Consider $y_2$ and
$y_3$, where the \pn they share would be the \pn in $y_2$ ($2n_p^2$),
plus the \pn in the first read, $p_{1,1}$.  Thus,
\begin{eqnarray}
p_{i,j} = p_{j,i} = kn_p^2 + p_{1,1},
\end{eqnarray} 
where
\begin{eqnarray}
k = \Big \lbrace {\begin{array}{cc}
 j, & j < i \\
 i, & i < j \\
\end{array}}.
\end{eqnarray}
If $p_{1,1}$ is included in the background, then it is set to zero.
We can estimate $n_p$ using the slope as we did in
Section~\ref{s:sim}: $n_p = \sqrt{s_e}$.  This initial calculation of
the slope for $s_e$ (eq.~\ref{eq:se}) is done before any \crr
correction and without taking into account correlated errors.  

This technique accounts for the diagonal elements as well.
Therefore, 
\begin{eqnarray}
  \begin{array}{lr}
    c_{i,j} = n_p^2  k + p_{1,1} + n_r^2 &  i = j, \\
    c_{i,j} = n_p^2  k + p_{1,1} & i \neq j, \\
  \end{array}
\end{eqnarray}
where $n_r$ is the \rn.  

An equation for the noise in a ramp has previously been derived in
nonmatrix form in~\citealp{Rau07}; for an independent derivation and
the correct final formula, see~\citep{Rob09_2}. \citeauthor{Rau07}
formula (eq. (1) in that article, see erratum) has the benefit that it
takes into account grouped samples.  However, the benefit of using
matrix notation with a covariance matrix is that we can add other
noise terms easily by calculating the appropriate covariance matrix
(like we did for \textbf{P} and \textbf{R}) and sum all to get
\textbf{C}.  It would be more difficult to add other noise terms to
the~\citet{Rau07} equation, and like our method here, they only
include \rn and \pn.

\subsection{Validating with Simulations}

\begin{figure*}
  \epsscale{1.1}
  \plottwo{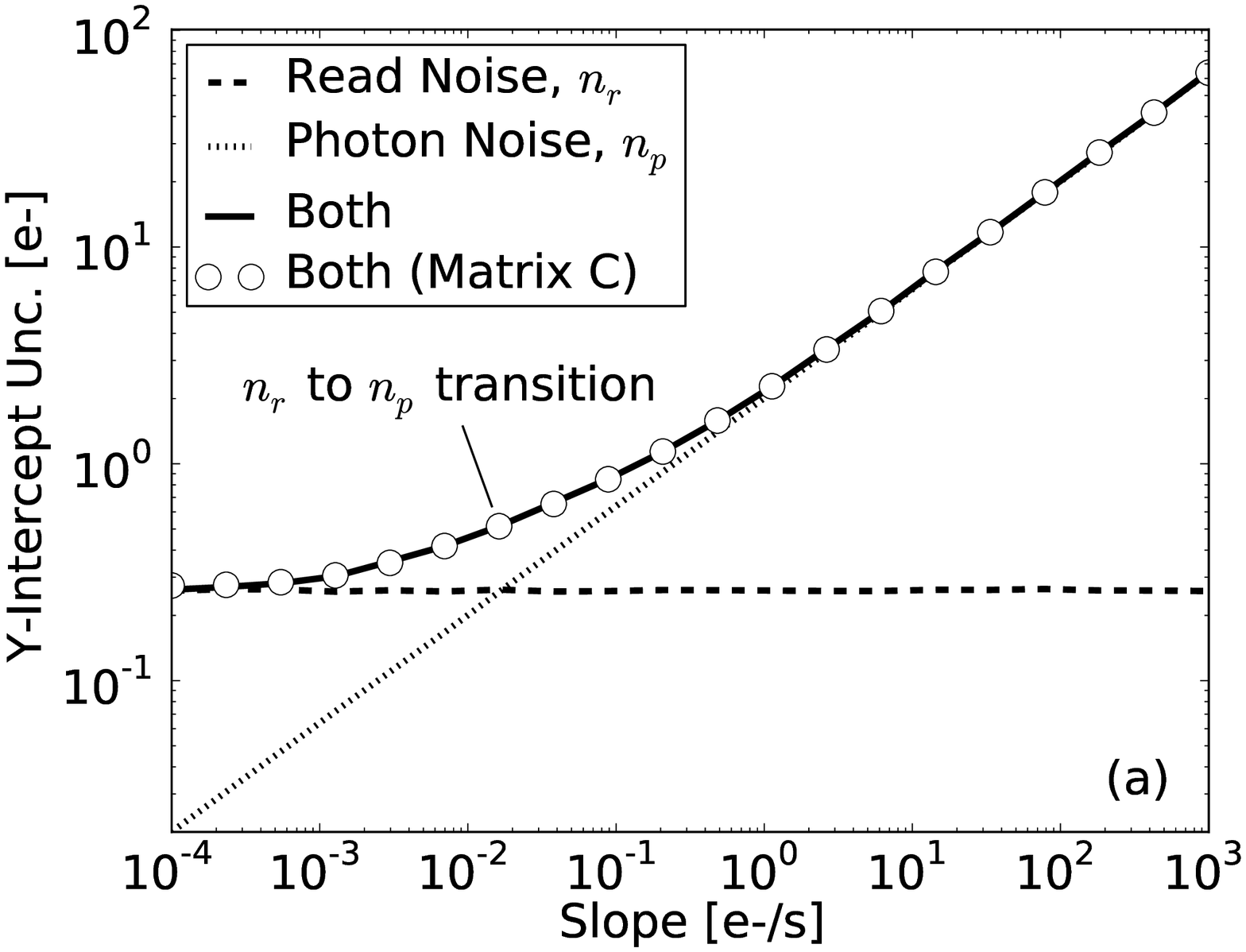}{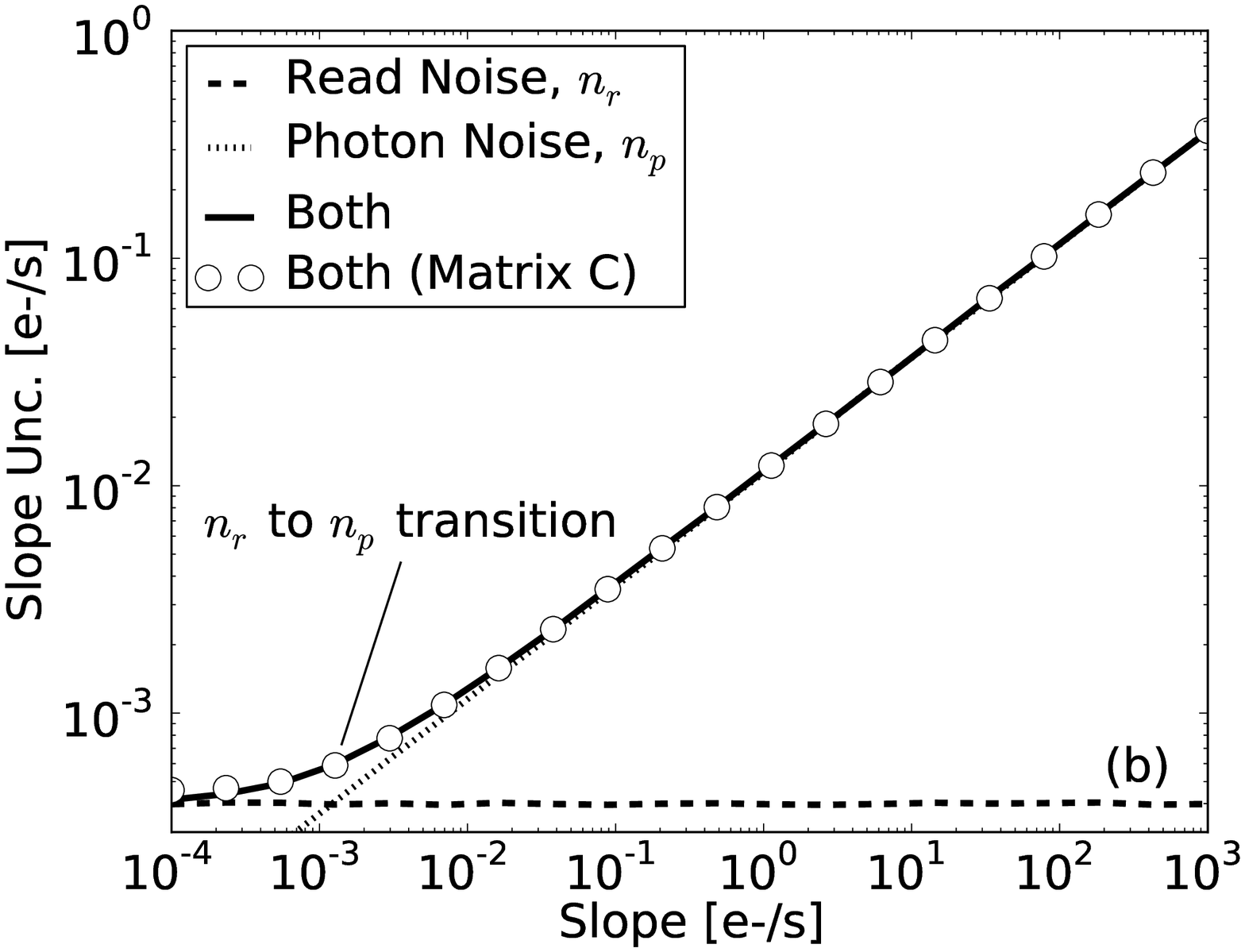}
  \caption{Calculation of the uncertainties
    for the y-intercept (a) and the slope (b) using the method
    described in Section~\ref{s:corr}.  Following the idea from
    Figure~16 in~\citet{Gor05}, for 10,000 each, we generated ramps
    with \rn only, \pn only, and then both.  The uncertainties in the
    y-intercept and slope given by the data are plotted by the dashed,
    dotted, and solid lines.  Then using correlated and uncorrelated
    components in the covariance matrix (e.g., including both \rn and
    \pn) we calculated the corresponding uncertainties shown by the
    circles.  These calculations match the data perfectly.
    Furthermore, this plot also shows where the transition is between
    the \rn-dominated regime and the \pn-dominated regime.}
  \label{fig:fitline_ys}
  \epsscale{1}
\end{figure*}

To demonstrate how well this calculation of the uncertainties fits
simulated data, we followed the idea from Figure~16 in \citet{Gor05}
and simulated 10,000 ramps each with \rn only, \pn only, and both,
then we used the covariance matrix to calculate the y-intercept and
slope uncertainties.  This is given in Figure~\ref{fig:fitline_ys}.
The dashed and dotted lines are the uncertainties with \rn and \pn
only, respectively.  The solid lines are the uncertainties in the
ramps where both \rn and \pn were added.  Notice that you can see
where the transition is between the \rn-dominated regime and the
\pn-dominated regime using these plots.  The circles are the
uncertainties calculated using the method described in
Section~\ref{s:corr}, which fit the data perfectly.

\begin{figure}
  \plotone{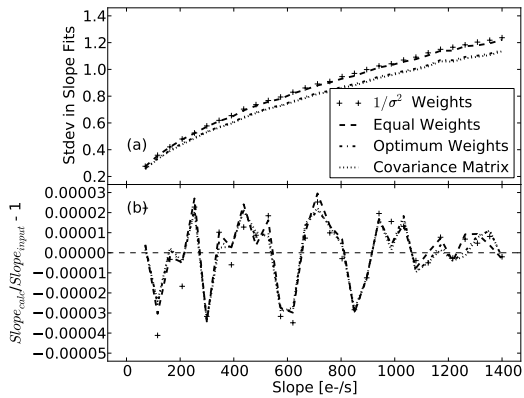}
  \caption{\textsl{(a) Standard deviation of slope fits versus input
      slope, for various weighting schemes to point out that using the
      covariance matrix has lower noise (though it is very similar to
      optimum weighting, as optimum weighting is an estimate of the
      covariance matrix).  (b) Calculated slope to input slope ratio
      minus one for the same methods.  All methods show the same bias;
      therefore, the difference lies only in the S/N shown in (a).}}
  \label{fig:linfit}
\end{figure}

In addition to using a covariance matrix, other popular methods to fitting a
line to ramps have used uncertainties as the weights (1/$\sigma^2$),
equal weights, or optimum weights, as discussed in~\citet{Fix00}.  To
demonstrate how these methods compare, we simulated 10,000 ramps, each
with the same input slope (before the noise was added) and then tried
to retrieve these slopes using the various weighting schemes
(Figure~\ref{fig:linfit}).  The standard deviation in the slope fits
is plotted in Figure~\ref{fig:linfit}~(a).  We see that the covariance
matrix and optimal weighting result in a higher signal-to-noise ratio (S/N) than
the other two methods.  Optimal weighting estimates a covariance
matrix, so it is expected that they show similar results.  In
Figure~\ref{fig:linfit}~(b) we show the ratio of the average
calculated slope to the input slope, with unity subtracted.  Note that
the curve in this plot is nonsmooth due to the finite number of
trials.  The error on the slope calculation is very small, with the
calculated slope at most 0.004\% from the input slope.  For various
slopes we made a histogram of the distribution of these ratios and
found that they were symmetric and centered on the average.
Therefore, optimal weighting and using a covariance matrix with
correlated and uncorrelated errors produces a lower standard deviation
of the results, while all methods produce a similar slope estimate
with 10,000 trials. 

\section{CR Detection Methods}\label{s:methods}
All three \crr detection methods have a unique algorithm.  However,
all follow the same procedure to calculate a slope once the \crs are
found, which we now outline:
\begin{enumerate}
\item Detect \crs, always looking for the largest outliers first.
\item Calculate the slope for the resulting ramp segments before and
  after the \crr events \citep[we will refer to these as ``semiramps''
  from here on, following][]{Rob08}.
\item Calculate the final slope of the entire ramp.  If there is one
  \crr or more, do this by taking the weighted average of the slopes
  of the semiramps (see also the discussion in Section~\ref{s:dis}).
\end{enumerate}

Furthermore, for all methods we use the absolute difference so that we
remove outliers in both directions and do not bias the data
\citep{Off99, Off01}.  However, if larger rejection thresholds are
used, and therefore there are no longer as many outliers (i.e., only
picking up the strongest \crs), then one-sided clipping would work
just as well~\citep{Win94}.

\subsection{Two-Point Difference Method}\label{ss:2pt}

With the nondestructive readout method, algorithms for \crr detection
include computing the two-point differences.  \citet{Off99} and
\citet{Fix00} describe versions of this method, though both use only
one rejection threshold for all cases (5$\sigma$ and 4.5$\sigma$,
respectively).  Another version was used in the \emph{Multiband
  Imaging Photometer} (\emph{MIPS}) data reduction algorithm for
\emph{Spitzer}~\citep{Gor05}.  We describe a variant of this
technique here.

\begin{figure}
  \plotone{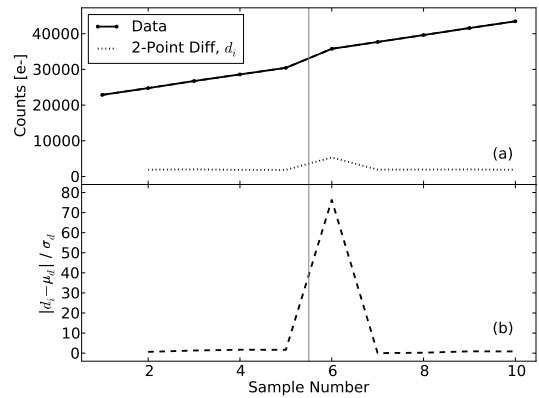}
  \caption{\textsl{\dif method.  (a) Solid
      line and points are the data, $y_i$, and the dotted line is the
      two-point differences, $d_i$, of this data.  (b) Dashed
      line is the ratio in equation~\ref{eq:2pt} which will be
      compared with the rejection threshold, $r_t$.  The vertical line
      shows the interval where the \crr hit.}}
  \label{fig:2pt}
\end{figure}

For the two-point difference method (hereafter \dif), we calculate the
two-point differences between the counts in each set of adjacent
samples.  The largest outlier is flagged as a \crr, given that it
fulfills the rejection criteria:
\begin{eqnarray}
\frac{\vert d_i -\mu_d \vert }{\sigma_d} > r_t,
\label{eq:2pt}
\end{eqnarray} 
where $d_i$ is the difference between the science data $y_{i+1}$ and
$y_i$, $\mu_d$ is the median of $d$, $\sigma_d$ is the
uncertainty of $d$, and $r_t$ is the rejection threshold.  The
median is used because it is more robust than the mean when there are
outliers in the data \citep{Pre86, Off99}.  The median can become a
problem if there is quantization noise (i.e., if the slope of the ramp
were 0.07 $e^-$/s), however we do not take into account quantization
noise in this article.  Once a \crr is identified, the $d_i$ that
includes the \crr is removed, and the process is repeated on the
remaining $d_i$ until no more \crs are found.  This method is
depicted in Figure~\ref{fig:2pt}.

When calculating the uncertainty in $d$, $\sigma_d$, the most obvious
solution would be to use the standard deviation.  However we have
found that this does not work well for ramps with a low number of
samples ($\sim5$). We can improve $\sigma_d$ by using the \pn and \rn
added in quadrature.  The \pn can be calculated as Poisson noise, but
since we are dealing with the two-point difference we use the charge
accumulated since the last sample, rather than the total charge in a
sample, so we use $\sqrt{s_e}$.  Since a \crr can contaminate a slope
calculation, we can use $\mu_d$ to estimate $s_e$.  Therefore $n_p =
\sqrt{\mu_d}$, and
\begin{eqnarray}
\sigma_d = \sqrt{n_p^2 + 2 n_r^2},
\label{eq:sigmad}
\end{eqnarray} 
where the factor 2 is due to the \rn from each of the two samples.

\subsection{Deviation from the Fit Method}\label{ss:dev}
The method we will refer to as the deviation from the fit method
(hereafter \dev), is the one used by \emph{NICMOS}~\citep{Dah08} and
\emph{WFC3}~\citep{Dre10}.  To use this method we fit a line to the
ramp using a covariance matrix as described in
Section~\ref{s:corr}. Then, for each sample, we calculate the
difference to the fit as a ratio to the uncertainty in the counts:
\begin{eqnarray}
dev_i = \frac{y_i - f_i}{\sigma_y},
\label{eq:diffn}
\end{eqnarray}  
where $f_i$ is the fit at sample $y_i$, and $\sigma_y$ is the
uncertainty in each sample, defined in equation~\ref{eq:yerr}.

We then take the first difference of these ratios, and look for the
largest.  If it satisfies the criteria:
\begin{eqnarray}
\vert dev_{i+1} - dev_i \vert  > r_t,
\label{eq:dev}
\end{eqnarray}  
then $y_{i+1}$ is flagged as a \crr.  The ramp is then split into
semiramps, and this method is applied again to the resulting
semiramps.

\begin{figure}
  \plotone{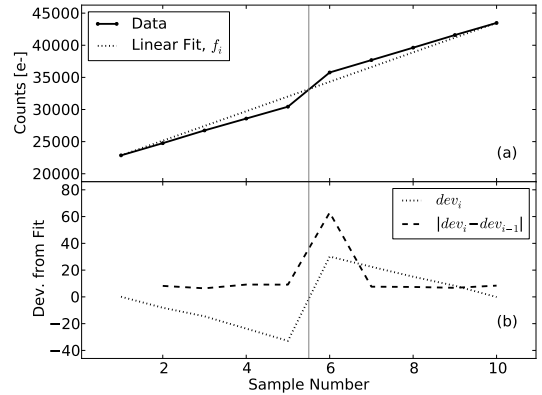}
  \caption{\textsl{\dev method.  The solid line
      and points are the data, $y_i$ and the dotted line is the linear
      fit made up of points, $f_i$.  In the lower panel the dotted
      line is the ratio of the difference between the fit and the data
      to the uncertainty, $dev_i$.  The dashed line is the two-point
      differences of the $dev_i$.  This is what will be compared with
      the rejection threshold, $r_t$.  The vertical line shows the
      interval where the \crr hit.}}
  \label{fig:dev}
\end{figure}

The \dev method is illustrated in Figure~\ref{fig:dev}.  Note that
the background level is not at zero as it was for the \dif method, but
$\sim10$.  The reason for the nonzero background is that while the
median (which would exclude the \crr) is subtracted in the \dif
method, in the \dev method the fit is subtracted, which includes the
\crr in its calculation.  Furthermore, in the presence of a \crr, the
slope will be overestimated and, therefore so will the \pn.  Finally,
the peak is at a ratio of $\sim$60, whereas both the \dif and \yint
methods (as we will see) peak at ratios of $\sim$80.

\subsection{Y-Intercept Method}\label{ss:yint}
\begin{figure}
  \epsscale{0.85}
  \plotone{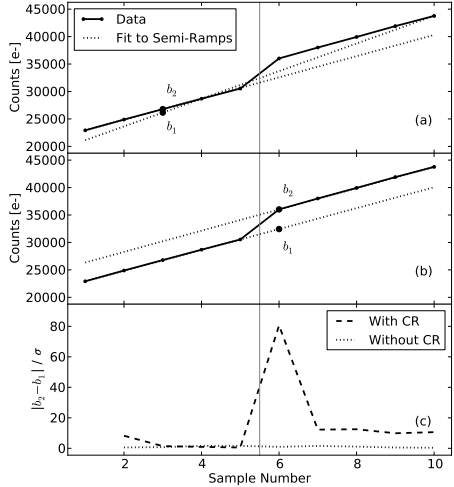}
  \caption{\textsl{\yint method.  (a) Counts and two linear fits as it
      would look when assuming that the \crr is in sample 3.  The
      solid line and points are the data, and the dotted lines are the
      linear fits to the semiramp before and after the assumed \crr.
      The two y-intercepts, $b_1$ and $b_2$ are highlighted with dots
      and labeled.  (b) Assuming the \crr is in sample 6 (which indeed
      it is).  (c) Results after stepping through the entire ramp
      assuming a \crr is in each sample.  Plotted is the ratio of the
      absolute difference between the y-intercepts to the uncertainty,
      which must be above the rejection threshold, $r_t$, to be
      counted as a \crr.  The interval where the \crr hit is shown by
      the vertical line.}}
  \label{fig:yint}
\end{figure}

The idea for the y-intercept method (hereafter \yint) comes from the
\emph{MIPS} data reduction algorithm~\citep{Gor05}.  The details we
describe here are a variant of that method.  For the \yint method, we
step through each sample and assume that there is a \crr there (the
first sample is skipped).  We fit a line to the semiramps before and
after the sample with the assumed CR, using a covariance matrix as
described in Section~\ref{s:corr}.  At each sample, the x-axis is
shifted so that the y-intercept is located at the sample number of the
assumed \crr.  The only exception is when we assume that there is a
\crr in the second sample.  Since we cannot calculate a y-intercept
for only one point, we shift the x-axis to the first sample, and use
the counts in that sample as the y-intercept, and take the \rn as the
uncertainty.  We do the mirror to this when assuming that there is a \crr
in the last sample.  Then, we take the ratio of the absolute
differences between the two y-intercepts ($b_1$ and $b_2$) to the
expected uncertainty, $\sigma_b$ (equation~\ref{eq:yint}).
\begin{eqnarray}
\frac{\vert b_2 - b_1 \vert }{\sigma_b} > r_t
\label{eq:yint}
\end{eqnarray} 
After stepping through every sample, we look at the sample with the
largest ratio.  If this ratio is larger than a given $r_t$, then we
flag that sample as a \crr.  The process is repeated on the resulting
semiramps.  The method is depicted in Figure~\ref{fig:yint}.  

The expected uncertainty, $\sigma_b$, is calculated from the
y-intercept uncertainty from the \rn and the y-intercept uncertainty
from the \pn added in quadrature.  There are two \rn components,
$n_{r_1}$ and $n_{r_2}$, one from the ramp before the assumed \crr,
and one from the ramp after.  These \rn terms are calculated by
setting the covariance matrix, $\bf{C}$, to the \rn matrix,
$\bf{R}$, and recalculating the y-intercept uncertainty.  The \pn
component is calculated as $n_p = \sqrt{s_e }$.  Since in the \yint
method we are dealing with two ramps, we take the weighted average of
the slopes to get $s_e$, just as we would if there were actually a
\crr found at the assumed location.  Therefore,
\begin{eqnarray}
\sigma_b = \sqrt{n_p^2 + {n_{r_1}}^2 + {n_{r_2}}^2}.
\end{eqnarray} 

In Figure~\ref{fig:yint}~(c) the dashed line representing
equation~\ref{eq:yint} is hovering around zero before the \crr
detection, and then afterward it increases to $\sim$10.  This effect
comes from the difference between the y-intercepts.  It is caused by the
fact that we shift the x-axis to the sample with the assumed \crr,
whereas if we shift it to the sample before the \crr (as we do for a
\crr in the second sample, notice that there the ratio is $\sim$10),
then the difference between $b_1$ and $b_2$ before the actual \crr
would be greater than after the actual \crr.

\section{Results}\label{s:results}

To compare the success of each method at finding \crs, we look
at both the fraction of \crs found and the false detection
rate, which we define as the ratio of the number of false detections to
the number of possible false detections.  The number of possible false
detections is calculated as the number of samples minus one,
minus the number of true \crs, times the number of trials.

We look at simulated ramps with 5, 10, 20, 30, and 40 samples, with
\crs of 20 different magnitudes from 0 to 250 $e^-$ (all \crs with
magnitudes $>$250 $e^-$ are found with all methods), and with slopes
of 0.0 $e^-$/s, 0.7 $e^-$/s, 3.5 $e^-$/s, 7.0 $e^-$/s, 35.0 $e^-$/s,
and 70.0 $e^-$/s, with one, two, or three \crs, and with the \crr located in
the beginning, middle, and end of the ramp (e.g. for a ramp with 40
samples, we inserted a \crr in the $3^{rd}$, $10^{th}$, $20^{th}$,
$30^{th}$, and the $35^{th}$ sample).  We then simulate 10,000 ramps
for each combination, and apply each method to each ramp to compare
the results. For all trials, the rejection threshold, $r_t$, is chosen
such that the rate of false detections is the same for all methods (a
rate of 5\% was chosen).  In this way we can best compare how well
they find the \crs.

\subsection{The Rejection Threshold, $r_t$}\label{ss:rt}
Each of the \crr detection methods has a different criterion (see
equations~\ref{eq:2pt},~\ref{eq:dev}, and~\ref{eq:yint}), which, when
compared with the rejection threshold, decides if a sample is to be
flagged as containing a \crr or not.  Therefore, the relationship
between rejection threshold and the false detection rate is different
for each method.  To better understand this dependence, for each \crr
rejection method we used a range of rejection thresholds and looked at
the resulting fraction of false detections.  This was done on \crr-free ramps.  

\begin{figure}
  \plotone{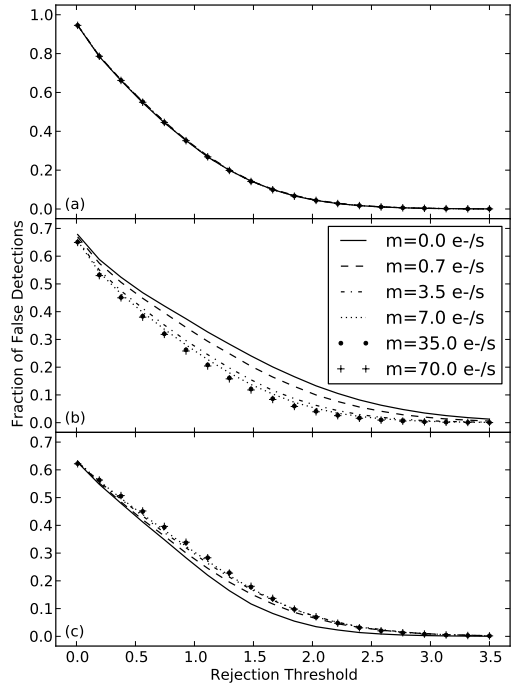}
  \caption{\textsl{Fraction of false detections versus rejection
      threshold is plotted here for various input slopes.  (a) \dif
      method.  (b)\dev method.  (c) \yint method.}}
  \label{fig:rt_slopes}
\end{figure}

The first comparison is for the fraction of false detections versus
the rejection threshold for different input slopes
(Figure~\ref{fig:rt_slopes}).  For a given fraction of false
detections, the rejection threshold is independent of slope for the
\dif method, while there is very little change for the \yint method.
However, for the \dev method, the rejection threshold needs to vary to
keep the fraction of false detections constant across ramp slopes.

\begin{figure}
  \plotone{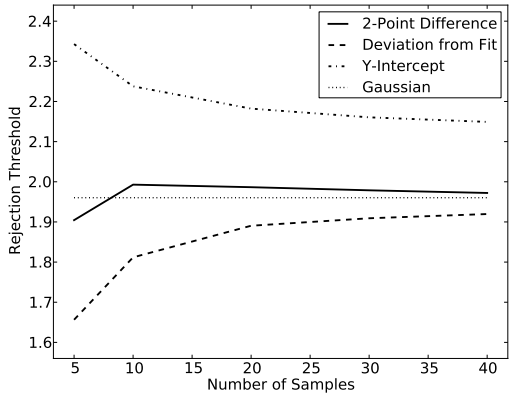}
  \caption{\textsl{Rejection threshold that will result in a
      fraction of false detections of 0.05 is shown for various sample
      numbers.  What we would expect from a Gaussian distribution is
      also plotted to show how similar it is to the \dif method.  A
      slope of 70.0 $e^-$/s was used for this figure.}}
  \label{fig:rt_nframes}
\end{figure}

The rejection threshold for different number of samples in a ramp is
shown in Figure~\ref{fig:rt_nframes}. An input slope of 70.0 $e^-$/s
was used, and the rejection threshold was chosen such that the
fraction of false detections was 0.05.  The slope of these lines is
0.0011, 0.0065, and -0.0049 for the \dif, \dev, and \yint methods
respectively.  Although this is a small change for the \dif method, if
a \crr is found in a ramp then the rejection threshold was changed
depending on the number of samples in a semi-ramp.  If the rejection
threshold was not calculated for a specific number of samples, we
interpolated it.  Figure~\ref{fig:rt_nframes} also shows how well the
uncertainties in the \dif method behave like a Gaussian, thus allowing
the rejection threshold to be chosen easily.

\subsection{Slope and CR Detection}\label{ss:slope}

\begin{figure*}
  \plotone{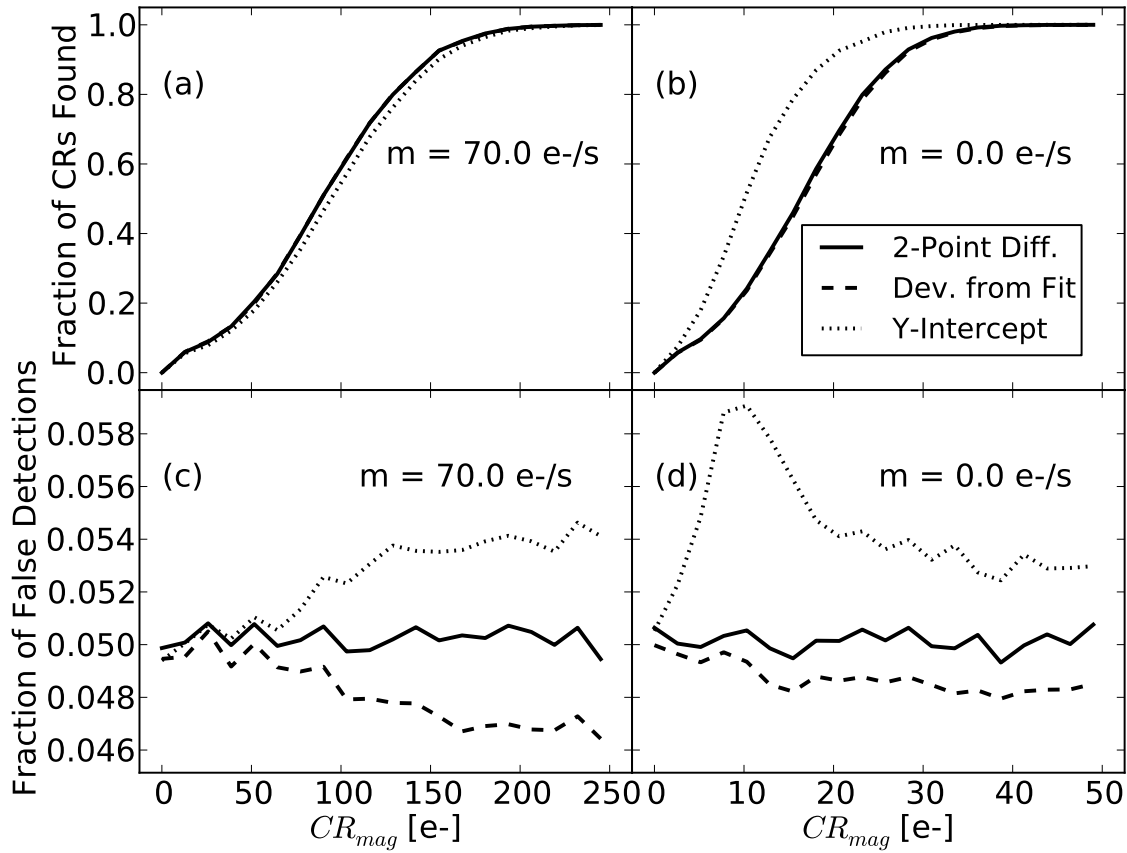}
  \caption{\crr detection rate and false detection
    rate as a function of \crr magnitude for all three methods.  In
    (a) and (c) the input slope was 70 $e^-$/s, in the \pn-dominated
    regime, where we recommend the \dif method be used.  In (b) and
    (d) the input slope was 0 $e^-$/s, in the \rn-dominated regime,
    where the \yint method outperformed the other two.  Each ramp has
    40 samples.}
  \label{fig:slope10_0}
\end{figure*}

As we change the input slope of these ramps, we move from the
\rn-dominated regime to the \pn-dominated regime.  In
Figure~\ref{fig:slope10_0} we show the two extremes: \rn-dominated
regime with an input slope of 0 $e^-$/s
(Figure~\ref{fig:slope10_0}~(b) and~(d)), and \pn-dominated regime
with an input slope of 70 $e^-$/s (Figure~\ref{fig:slope10_0}~(a)
and~(c)).  Plotted is the fraction of \crs found versus the \crr
magnitude and the fraction of false detections versus the \crr magnitude.

In the \pn-dominated regime, \pn can appear as a \crr as it is
correlated and thus can lead to false detections.  From
Figure~\ref{fig:slope10_0}~(a) and~(c) we see that all of the methods
give similar results.  However, note how the fraction of false
detections for the \dif method is relatively constant compared with
the other two methods.  This shows how well we are able to understand
the uncertainties of this method.  This, coupled with the fact that
the \dif method requires the least computations, leads us to suggest
that the \dif method is the optimal \crr detection method in the
\pn-dominated regime.

In the \rn-dominated regime we only have random noise, and therefore
we are able to get a more accurate calculation of the y-intercept and
the y-intercept uncertainties.  Therefore, as is shown in
Figure~\ref{fig:slope10_0}~(b) and~(d), in the \rn-dominated regime
the \yint method gives the best results in that it is able to find
fainter \crs than the other two methods.

In Figure~\ref{fig:slope10_0}~(d), note that the rate of false
detections per \crr strength is only constant for the \dif method,
while for the \yint method it is constant only for high-energy \crs.
Between a $CR_{mag}$ of 0 $e^-$ and 10 $e^-$, we see a high fraction
of false detections for the \yint method.  This corresponds to the
range where the \crs are not always found.  The cause of this 'bump'
is not clear, and is evidence that we do not fully understand the
noise model associated with the Y-INT method.

This could be caused by the fact that we treat each semiramp
independently, which is inaccurate, since the uncertainties are
correlated.  Therefore, one solution could be to fit all semiramps
simultaneously.  This would increase the uncertainty, which would mean
that the rejection threshold has been set too low (which would agree
with what we see in the plots).  We leave that for future work.

\begin{figure}
  \plotone{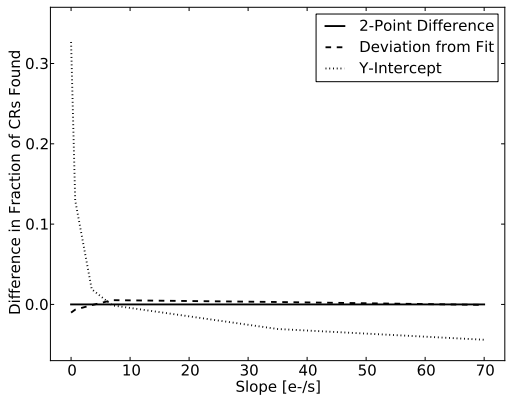}
  \caption{Difference between the fraction of \crs found by the \dev
    and \yint methods to the \dif method.  From this plot we are able
    to see that the \yint method outperforms the other two with slopes
    less than $\sim5 e^-$/s.}
  \label{fig:yint_vs_2ptdiff}
\end{figure}

Regarding Figure~\ref{fig:slope10_0} (a) and (b), if you draw a
vertical line at a given $CR_{mag}$ (we chose the $CR_{mag}$ where the
fraction of \crs found by the \dif method is closest to 50\%), you can see
the difference between the fraction of \crs found by each method at
that slope.  We did this for slopes of 0.0$e^-$/s, 0.7$e^-$/s,
3.5$e^-$/s, 7.0$e^-$/s, 35.0$e^-$/s, and 70.0$e^-$/s.  The results are
shown in Figure~\ref{fig:yint_vs_2ptdiff}.  Each ramp had 40 samples,
and again the rejection threshold was chosen such that the fraction of
false detections was 5\%.  For comparison, we subtracted the fraction
of \crs found by the \dif method.  In this figure we can see that we
move into the regime where the \yint method performs better than the
\dif and \dev methods at a slope of $\sim5 e^-$/s.

\begin{figure}
  \plotone{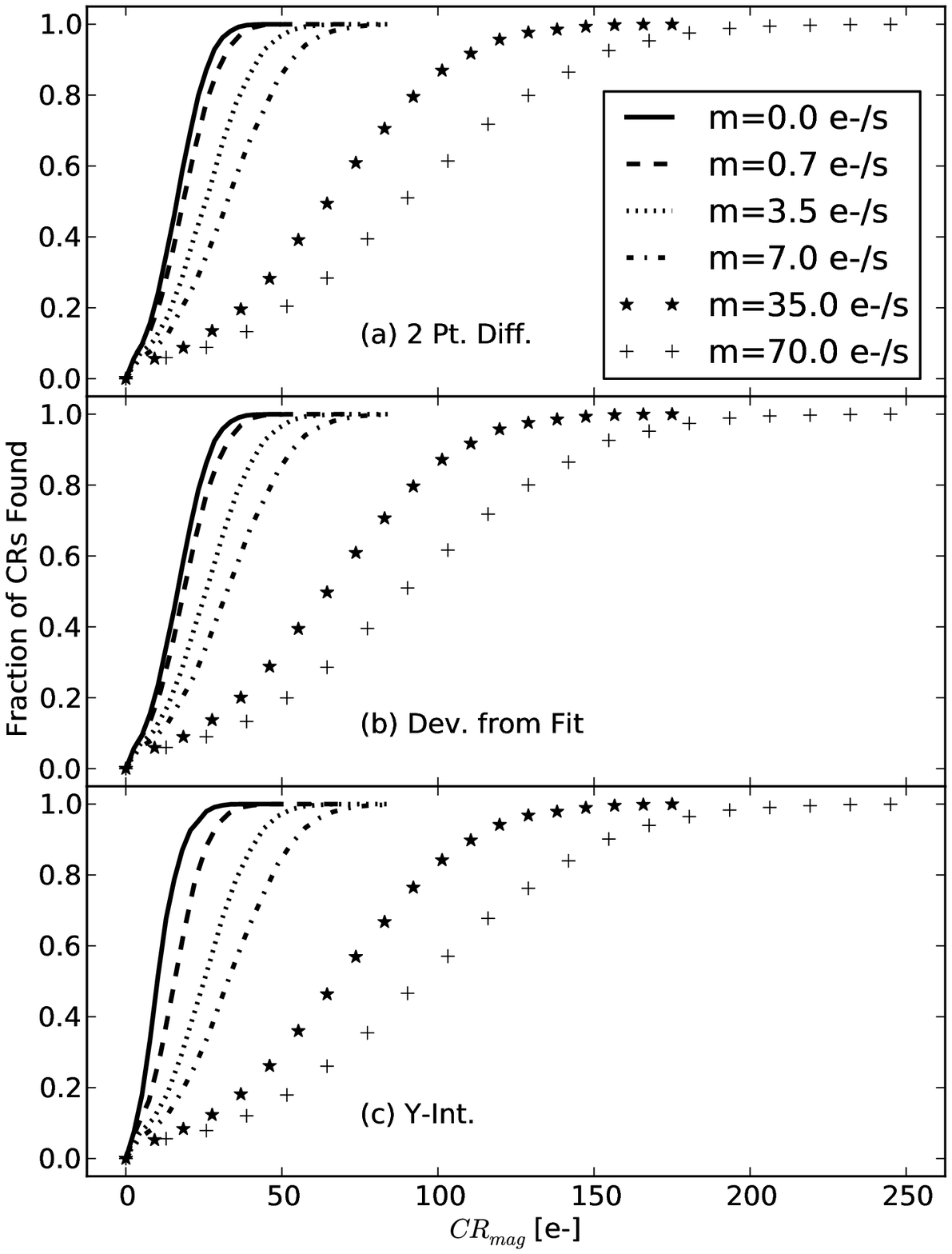}
  \caption{\crr detection rate as a function of \crr magnitude for all
    methods.  Each line is for a different input slope.  There are 40
    samples in each ramp.}
  \label{fig:slopes}
\end{figure}

To further illustrate how the results of each method change with input
slope, we show the fraction of \crs found as a function of \crr
magnitude for different slopes and for each of our methods in
Figure~\ref{fig:slopes}.  The results of the \dif and \dev methods
appear to be similar, while the \yint method does better with lower
slopes (\rn-dominated regime).

\subsection{Number of Samples and CR Detection}\label{ss:nframes}
\begin{figure}
  \plotone{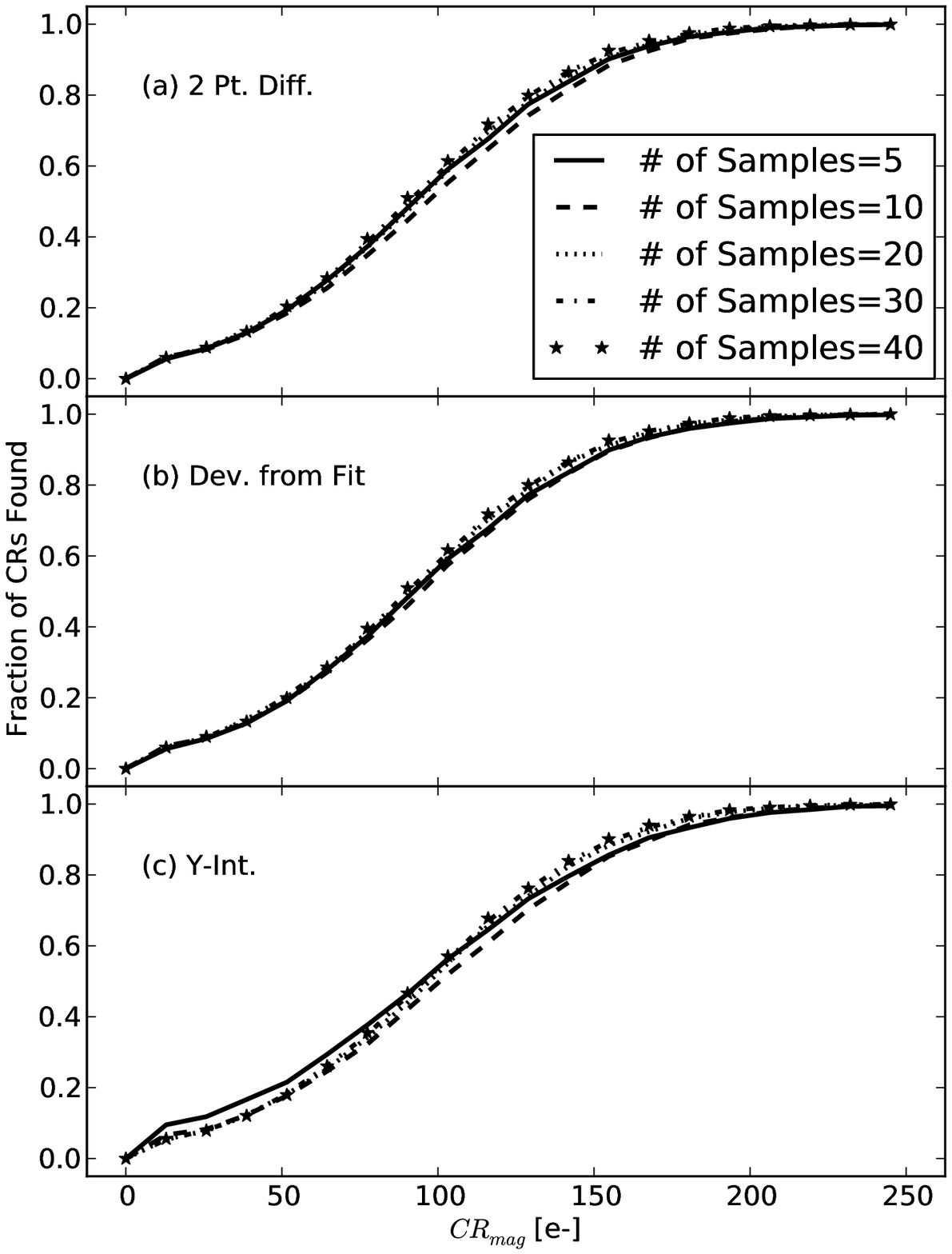}
  \caption{\crr detection rate as a function of
    \crr magnitude for all methods.  Each line is for a different
    number of samples in a ramp.  The input slope is 70 $e^-$/s for each
    ramp.}
  \label{fig:nframes}
\end{figure}

In order to determine how the number of samples in a ramp affects how
well we detect \crs, we tested each algorithm on simulated ramps with
5, 10, 20, 30, and 40 samples.  Figure~\ref{fig:nframes} shows the
fraction of \crs found as a function of \crr magnitude for different
numbers of samples for the \dif, \dev, and \yint methods.

When we use the \dif method on weak \crs ($<$ 50 e$^-$) and strong
\crs ($>$ 150 e$^-$), the fraction of \crs found increases with the
number of samples (except for the case where there are five samples).
For the \dev method, the fraction of \crs found increases with number
of samples for string \crs, but shows no difference for weak \crs.
Finally, for the \yint method, the fraction of \crs found decreases
with number of samples for weak \crs, but increases with number of
samples for strong \crs.  

\subsection{CR Sample Number and CR Detection}\label{ss:framen}
There are two questions when it comes to \crr sample number: the first
is where do we find false detections in the ramp, and the second is
how well we find \crs in given positions in the ramp.

\begin{figure}
  \plotone{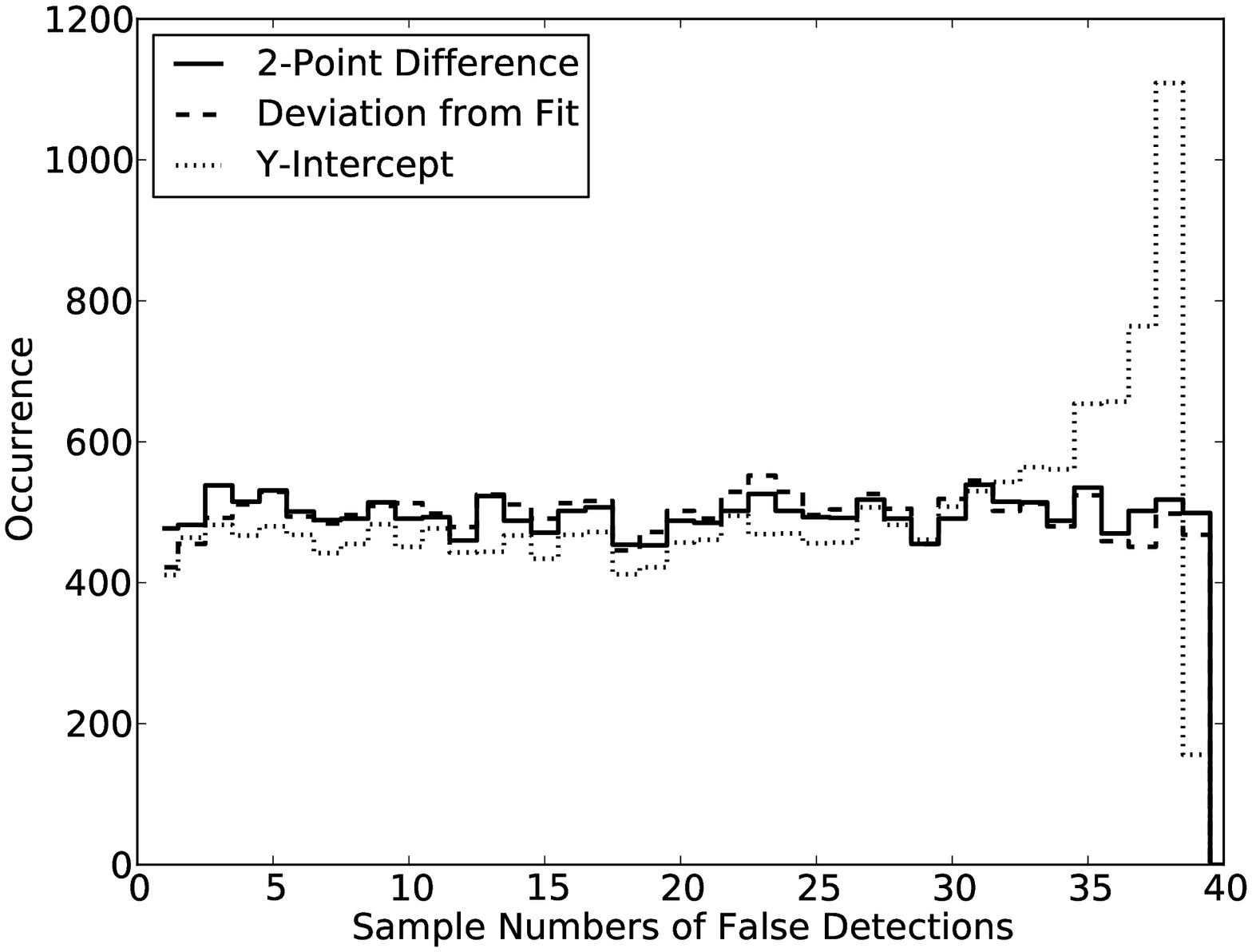}
  \caption{\textsl{Histograms of the sample numbers of
      the false detections for each method.  These are from ramps
      without a simulated \crr.}}\label{fig:rn4false}
\end{figure}

To answer the first question, we created a histogram of the sample
numbers of the false detections found on simulated \crr-free ramps for
each method.  These are presented in Figure~\ref{fig:rn4false}.  The
results show that both the \dif and \dev methods are not biased
toward any position in the ramp.  However, the \yint method is biased
toward the last samples in a ramp where there is less effect from the
correlation between samples.

\begin{figure}
  \plotone{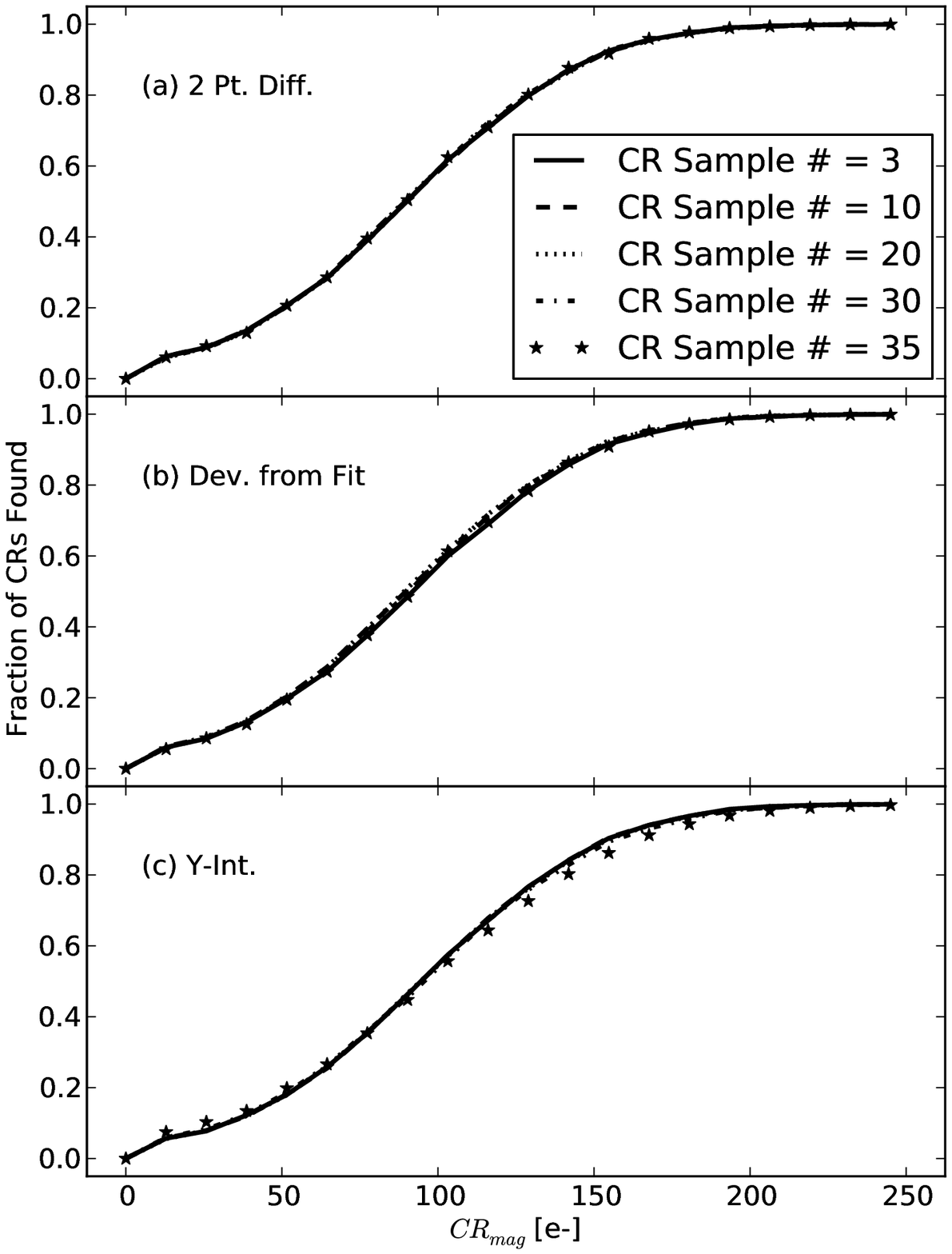}
  \caption{\crr detection rate as a function of \crr magnitude for all
    methods.  Each line is for a different \crr sample number.  The
    input slope is 70 $e^-$/s, and there are 40 samples in each ramp.
    The \dif method is the only one to not show a dependence on sample
    number.}
  \label{fig:readn}
\end{figure}

To answer the second question, we simulated ramps with 40 samples and
input slope of 70 $e^-$/s, with one \crr each at sample (numbers 3, 10, 20,
30, or 35).  We then applied each of the algorithms to these ramps and
compared the results, shown in Figure~\ref{fig:readn}.

The results of these plots all show that the \dif is the only method
that does not vary depending on the sample number of the \crr, as can
be expected.  Both the \dev and \yint methods are weakly biased by
sample number.  This is also expected, since both require fitting
lines to data, and the fewer points in a line the less accurate the
fit.  Again, in the \pn-dominated regime, the \dif method is best.

\subsection{Multiple Cosmic Rays and CR Detection}\label{ss:multi}

In a 1000 s integration, if we assume that 20\% of the field
will be affected by \crs, then about 4\% of the field could be
affected by two \crs, and 0.8\% could be affected by three.  This is
substantial enough that we need to account for this possibility.

\begin{figure}
  \plotone{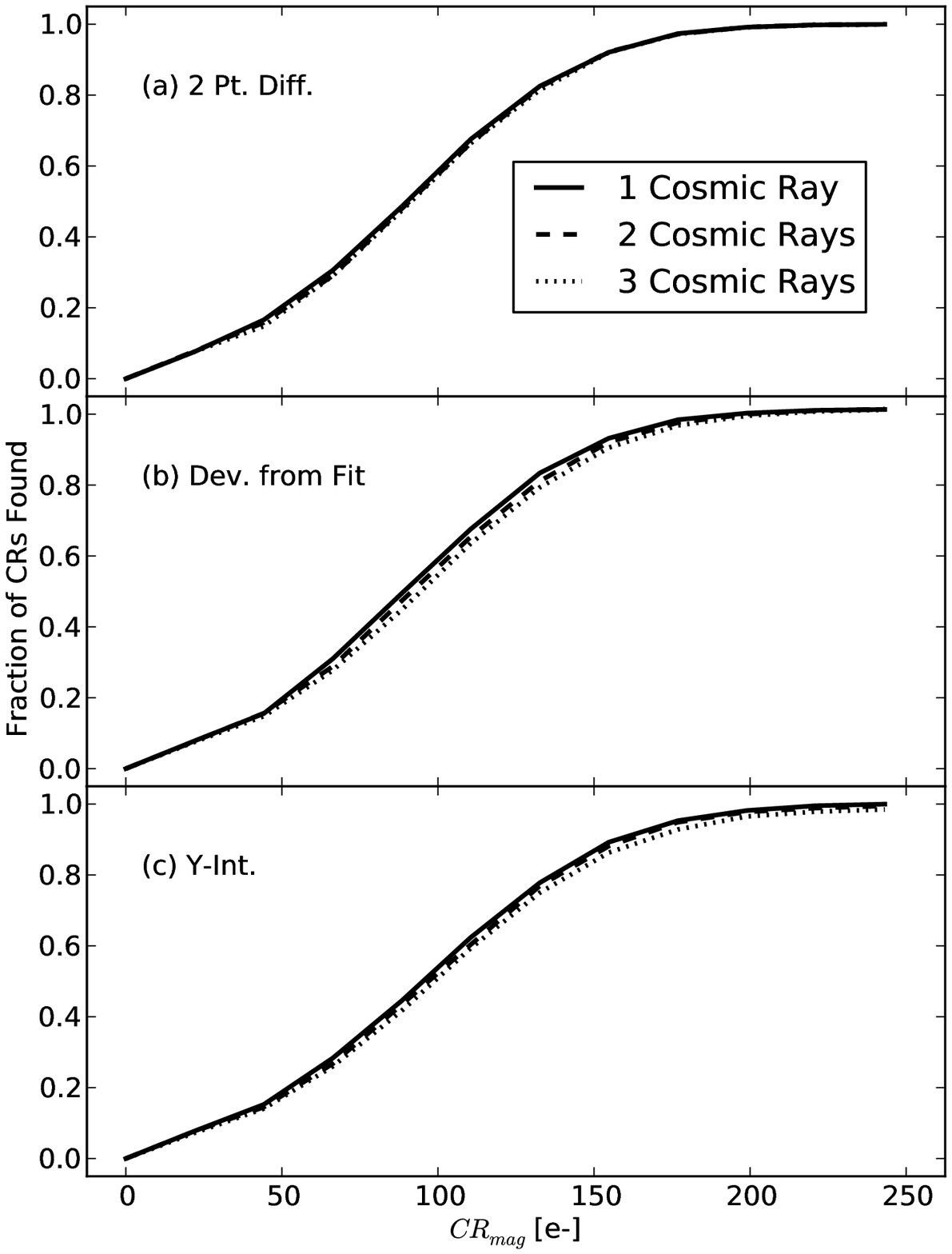}
  \caption{\crr detection rate as a function of \crr magnitude.  Each
    line is for a different number of \crs in a ramp.}
  \label{fig:mCRs}
\end{figure}

To test how well each method does at handling multiple \crs, we
simulated 10,000 40-sample ramps for each of 0-19 \crs.  Each \crr was
of the same strength and in random but different samples. We show the
results for 0-3 \crs in Figure~\ref{fig:mCRs}.

As can be seen, the \dif method shows no difference between one, two,
or three \crs.  We show only up to three \crs for simplicity, but we
have found that the \dif method does not start to show a difference in
results until about nine \crs and therefore should be the optimal method
in the \pn-dominated regime.  Both the \dev and \yint methods,
although show more variance than the \dif method.  This is caused by
the fact that both methods require linear fits that will include two
\crs for the \yint method (when assuming the correct location of one
of the \crs), and all three \crs for the \dev method.  Furthermore, for
both methods the ramp will be segmented after the detection of the
first \crr.


\section{Discussion}\label{s:dis}

The three \crr detection methods presented in this article were tested
on simulated ramps, adjusted, corrected, and refined, in order to
present optimum versions.  We now compare our methods to determine
which is best suited for various data.

\subsection{2-Point Difference Method}

Our results show that the \dif method performs best in the
\pn-dominated regime.  On top of being straightforward and
computationally simple, we showed that the fraction of false
detections versus rejection threshold is consistent regardless of
slope and number of samples.  We also showed that the uncertainties of
the \dif method follow a Gaussian distribution.  Any deviation from a
Gaussian was initially thought to be caused by the use of the median
instead of average in the rejection criterion.  However we found that
while using the average did produce a more Gaussian shape, it was
still not a perfect Gaussian.  It was demonstrated that, as expected,
the fraction of \crs found changed with slope and number of samples
(though barely).  Furthermore, we showed that the false detections are
evenly distributed in all samples, and that the \crr sample number did
not change the results.  Finally, we found that there is no noticeable
difference in the performance of the \dif method with multiple \crs,
up to ~9 \crs on a ramp with 40 samples.

Overall, the \dif method is fast, consistent, and easy to understand and
calibrate (e.g., choose a rejection threshold), and it gives the best
results in the \pn-dominated regime.

\subsection{Y-Intercept Method}

For the \yint method, there are two regimes:
\begin{enumerate}
\item Photon-noise-dominated: In this regime the \yint method gives
  the same answer as the \dif method.  This is due to the correlated
  behavior of the noise in the ramps (e.g., a $3\sigma$ event due to
  the noise of the photons that have arrived in the last sample time
  will propagate through all subsequent samples, having the same
  effect as a \crr).  Calculating the linear fit of subsequent samples
  will not negate the fact that the noise in the photon-dominated
  regime is set by the noise in a single sample.
\item Read noise-dominated: In this regime the \yint method is better
  than the \dif method.  These two methods will still give the same
  results on any noise above the \dif threshold, but anything below
  that will only be detectable by the \yint method.  With the noise
  independent from sample to sample, a linear fit reduces the
  uncertainty in the y-intercepts and weaker \crs are detectable.
\end{enumerate}

While it is unlikely that the uncertainty will be \rn-dominated for
most \emph{MIRI} data (given the telescope and sky background), it
will be the case for the \jw \emph{NIRSpec}, \emph{NIRCam}, and the
\emph{Tunable Filter Imager} (\emph{TFI}).  \emph{NIRSpec} has a
higher resolution than the \emph{TFI} and therefore a lower
background.  Regardless, the \emph{TFI} will likely be \rn limited,
given that it uses a tunable filter (narrowbands are less
sky/telescope-dominated).

The \yint method, however, does not quite match the robustness of the
\dif method.  In theory, these two methods should be the same in the
\pn-dominated regime.  However, the \dif method gave more consistent
results for different slopes, number of samples, and \crr location in
the ramps and for multiple \crs.  The difference is that the \dif
method has a simpler noise behavior, while we have not fully solved for
the noise behavior for the \yint method.  Despite being careful when
dealing with correlated and uncorrelated errors when fitting a line to
a ramp or semiramp, we did not take this into consideration when
taking the average of the slopes of the semiramps.  Instead we use a
weighted average, where the weights are simply the uncertainties in
the slope.  A solution to this problem would be to fit each semiramp
and the \crr simultaneously, as done by \citet{Rob08} for uncorrelated
errors only, modifying the method to account for correlated errors.
Regardless, when working in the \rn-dominated regime, the \yint method
should be used.

\subsection{Deviation from Fit Method}

The \dev method did not perform as well as the other two methods.  The
\yint method dominated in the \rn regime, and in the \pn-dominated
regime the \dev method, like the \yint method, was not as consistent
as the \dif method when it came to different slopes, number of
samples, \crr location in the ramps, and multiple \crs.  Therefore, to
have a consistent fraction of false detections, we would have to set a
new rejection threshold based on these variables.  Furthermore, while
not as complex at the \yint method, the \dev method is still more
complex than the \dif method and thus computationally more expensive.
Finally, the uncertainties are nowhere near Gaussian, thus making it
difficult to set the rejection threshold.  The \dev method was only
slightly worse at detecting \crs than the \dif method, but given the
fact that it is not as robust, and requires more CPU time, the \dev
method is not recommended.

The \dev method could be improved by changing the weighting scheme so
that instead of using the slope in the calculation of the \pn (since
the slope will still include the \crr in its calculation), we could
use the median of the two-point differences as we do for the \dif
method.  However, it still would not compare with the \dif method, as
we cannot do better than two$n_r$ terms and one $n_p$ term by the
nature of our problem.  Furthermore, even if we were able to get the
same results for these two methods, the \dev method still requires
larger computational resources than the \dif method.

\section{Conclusions}\label{s:cons}

In this article we discussed three \crr detection methods for
nondestructive read ramps, as well as their strengths and weaknesses.
We applied these methods to simulated ramps with single-sample groups.
We showed that the covariance matrix must include correlated errors in
order to improve the calculation of the slope and y-intercept of a
ramp and their uncertainties.  The \yint method benefits most from
this use of the covariance matrix, and would not be an improvement of
the \dif method otherwise.

The \dif method was shown to be able to give the best results in the
\pn-dominated regime.  The method's uncertainties are quasi-Gaussian
which simplifies the process of choosing a rejection threshold.
Moreover, it is fast and consistent.  Its robustness, compared with
the other two methods, resides in the fact that we simply remove the
two-point difference that includes the \crr, instead of splitting the
ramp into two semiramps, when searching for other \crs and calculating
the slope and the y-intercept.

The \dev method results in a similar fraction of \crs found as the
\dif method for some ramps, but unlike the \dif method these results
change based on slope, number of samples, and \crr location in the
ramp and for multiple \crs.  If we also consider the fact that it is
computationally more expensive, we do not recommend this method for
use in any regime.

The \yint method achieves the best results in the \rn-dominated
regime, and returns the same results as the \dif method in
\pn-dominated regime.  In the \pn-dominated regime, the \yint method
behaves like the \dif method in that it is effectively calculating an
average slope, excluding disturbances from any \crs (in the diff
method this is done by taking the median of the two-point differences).
The average slope is then divided by the noise due to one sample time:
one \pn and two \rn.  The \yint method is better in the \rn-dominated
regime as the uncertainties on the line fit parameters diminish in
size as more points are fit.  This is not the case in the
\pn-dominated regime due to the correlated nature of the noise.  The
\yint method only has two drawbacks: the noise model is not fully
understood, and it is a complex algorithm.

In summary, if we take all results into consideration we are led to
suggest that if only one method can be used on data cubes, the \yint
method should be it.  If computational speed is an issue then the \dif
method should be used, especially where \pn dominates, and the \yint
method should only be used when the estimated slope is in the
\rn-dominated regime or where there is information, such as a nearby
\crr and cross-talk is suspected, to look for fainter effects.


\section{Acknowledgments}\label{s:acknowledgments}
This work was supported by the Space Telescope Science Institute,
which is operated by AURA, Inc., under NASA Contract No. NAS5-26555.

We thank Howard Bushouse for providing the information on the method
used by NICMOS and WFC3, which was called the \dev method here.

Appreciation goes to Harry Ferguson, David Grumm, Derck Massa, Mike
Regan, and Massimo Robberto for their comments.



\clearpage


\clearpage






\end{document}